\documentclass[aps,pra,epsfigure,twocolumn,showpacs]{revtex4}
\usepackage{dcolumn}
\usepackage{bm,enumerate}
\usepackage{graphicx}
\usepackage{amsmath}
\usepackage{latexsym}
\usepackage{amsfonts}
\usepackage{amssymb}
\usepackage{array}
\usepackage{epsfig}
\usepackage{txfonts}

\newcommand{\ket}[1]{\left\vert#1\right\rangle}
\newcommand{\bra}[1]{\left\langle#1\right\vert}

\begin{document}
\title{%Sexiness?:\\
Predominance of entanglement of formation over quantum discord under quantum channels}
\author{Steve Campbell}
\affiliation{
Quantum Systems Unit, Okinawa Institute of Science and Technology, Okinawa 904-0495, Japan\\
Department of Physics, University College Cork, Republic of Ireland}
%ABSTRACT
\begin{abstract}
We present a study of the behavior of two different figures of merit for quantum correlations, entanglement of formation and quantum discord, under quantum channels showing how the former can, counterintuitively, be more resilient to such environments spoiling effects. By exploiting strict conservation relations between the two measures and imposing necessary constraints on the initial conditions we are able to explicitly show this predominance is related to build-up of the system-environment correlations.
\end{abstract}
\date{\today}
\maketitle

\section{Introduction}
%{\bf Needs polished}
The understanding of quantum correlations has come a long way since Schr\"odingers first mention of entanglement~\cite{schrod}. It is well established the potential advantages that quantum correlated systems offer over classical counterparts. Despite a substantial understanding of bipartite entanglement, and more generally non-classicality, we still lack a fundamental description of quantum correlations. Within a particular figure of merit's description there are ordering problems associated with mixed states~\cite{ordering,orderingdiscord}, while means to quantify and classify multipartite correlations are still in the early stages of being developed~\cite{giorgimulti,sarandy,jeremie}.  Quantum correlations are at the core of quantum information protocols, and more recently they have been linked explicitly to quantum phase transitions~\cite{QPT} and even in how biological systems behave~\cite{GOOLD?}. As it becomes clearer that quantum correlations are central to how physical systems behave, a complete understanding of them is rightly of great importance.

When addressing the issues arising when dealing with mixed states, one is implicitly asking the question of how quantum systems interact with their environments~\cite{breuer,maziero,vedral}. Given the fragility of a quantum state, and the impossibility of realizing ideal unitary conditions, understanding the dynamics of quantum correlations in open systems is pivotal. Until the discovery of states that do not violate a Bell inequality, and yet are inseparable, non-locality and entanglement where thought of as the same resource. Now with the plethora of means to assess the quantum nature of a state~\cite{horodecki,paterek} we are faced with a bigger challenge. 

In more recent years the study has significantly broadened from the well established notion of quantum entanglement to include a wider description aimed at quantifying the genuine non classicality  of a system. While theoretical advances have largely explored these concepts independently more recent studies have asked how they relate to each other in relevant physical settings, for example dynamics under local channels~\cite{bruss,francesco} and propagation along spin chains~\cite{campbell}. In many of these studies quantum discord~\cite{discord1,discord2} (QD) has presented itself as the more attractive potential resource. One conclusive difference between entanglement and QD shown was the latter decays asymptotically under Markovian-type noise for a certain class of states, while we would see the onset of entanglement sudden death~\cite{ESD}. Recently interesting works exploring the interplay between entanglement and discord~\cite{adesso,fanchini,campbell} have similarly shown discord serving as the more optimal resource. 
As explicitly shown in~\cite{marco} if one considers the correlations created between a system and some measuring apparatus by local measurements then the more general quantumness of the correlations are always greater than or equal to the entanglement, within such a framework. 
This leads to a natural assumption that the more general picture for quantumness, as captured by QD, is more resilient to noise than entanglement.  We show, however, this is not always the case and explore conditions under which the different notions of quantumness can be contrasted both qualitatively and quantitatively. 

Our study will explicitly address amplitude and phase damping environments, two widely applicable forms of noisy quantum processes, although the validity of the results can go beyond these examples. Using quantifiers with the same entropic definitions and ensuring initially the measures coincide we are able to explicitly show that entanglement of formation~\cite{wootters} (EoF) can, counterintuitively, be shown to be more resilient to the environmental action than QD. By exploiting a strict conservation relation between the two measures~\cite{fanchini} we are able to show how this is related to the dynamics of the correlations developed between the system and environment. 

The remainder of the paper is organized as follows. In Sec.~\ref{preliminaries} we introduce the tools necessary for a complete understanding of the work. Sec.~\ref{phasedamping} explores what happens under the action of phase damping environments. Sec.~\ref{ampdamping} covers the case of amplitude damping environments. In Sec.~\ref{mixedstates} we address the case of mixed initial states. Finally Sec.~\ref{conclusions} we summarize our conclusions and give some discussions on the results.

\section{Preliminaries}
\label{preliminaries}
Open quantum systems have attracted considerable theoretical study and as such there is a significant body of tools for us to draw from~\cite{breuer}. The study of bipartite quantum correlations similarly has amassed a number of quantifiers one could exploit depending on the specific questions to be addressed~\cite{paterek,horodecki}. As such, we will restrict ourselves to exploring two different types of damping that embody a wide range of physical settings and utilize measures that allow for reasonable comparison on the grounds of operational construction. We will first consider phase damping environments. Phase damping describes the loss of coherence while keeping the energy constant. It can be described by the single qubit Kraus operators~\cite{chuang} 
\begin{equation}
\label{phase}
K_0=\left(
\begin{array}{cc}
 1 & 0 \\
 0 & \sqrt{1-\lambda }
\end{array}
\right)~~~~~~~
K_1=\left(
\begin{array}{cc}
 0 & 0\\
 0 & \sqrt{\lambda} 
\end{array}
\right),
\end{equation}
where $0\leq\lambda\leq1$. Since the phase damping environment only affects the coherences, it will provide an informative platform to comparatively study different figures of merit for quantum correlations. We next consider amplitude damping environments. This describes the probability of losing an excitation to the environment and thus affects both the coherence and energy within a system. It can be described via the single qubit Kraus operators~\cite{chuang}
\begin{equation}
\label{amp}
K_0=\left(
\begin{array}{cc}
 1 & 0 \\
 0 & \sqrt{1-\gamma }
\end{array}
\right)~~~~~~~
K_1=\left(
\begin{array}{cc}
 0 & \sqrt{\gamma } \\
 0 & 0
\end{array}
\right),
\end{equation}
where $0\leq\gamma\leq1$. The action of the environments on a two qubit state can then be calculated 
\begin{equation}
\varrho=\sum_{i,j=0}^1(K_i\otimes K_j) \rho (K_i\otimes K_j)^{\dagger}.
\end{equation}
One can readily determine an extremal asymmetric damping scenario, where only a single qubit is damped while the other remains completely unaffected, by simply applying the identity instead of the Kraus operators to one of the qubits.  Both quantum channels are Markovian (memoryless), such a consideration allows us to clearly examine how the decaying dynamics of correlations behave. As we are only interested in exploring how the different figures of merit are adversely affected by decoherence the Markovian assumption is well justified here.  While it would be interesting to address non-Markovian environments, it is outside the scope of this work.

In order to assess how different figures of merit behave under such channels great care must be taken both at the level of initial states and correlation quantifiers. We consider the entanglement of formation (EoF) and quantum discord (QD), due to their construction they share the same entropic definition. The relationship between EoF and QD has been recently examined to study the distribution of quantum
correlated states~\cite{alqasimi,campbell,fanchini}. Moreover, it is possible to establish a strict  relationship connecting QD and EoF~\cite{fanchini}, so that such two figures of merit appear to be natural choices for a quantitative comparison

We take QD~\cite{discord1,discord2,paternostro} as a measure for general quantum correlations between any two qubits under study. As originally proposed by Ollivier and Zurek, QD can be associated with the difference between two classically equivalent versions of mutual information, which measures the total correlations within a quantum state. For a two-qubit state $\rho_{AB}$, the mutual information is defined as 
\begin{equation}
{\cal I}(\rho_{AB})={\cal S}(\rho_A)+{\cal S}(\rho_B)-{\cal S}(\rho_{AB}). 
\end{equation}
Here, ${\cal S}(\rho)\,{=}\,{-}\text{Tr}[\rho\log_2\rho]$ is the von Neumann entropy of the generic state $\rho$. Alternatively, one can consider the one-way classical correlation~\cite{discord2} 
\begin{equation}
{\cal J}^\leftarrow(\rho_{AB})={\cal S}(\rho_{A})-{\cal H}_{\{\hat\Pi_i\}}(A|B), 
\end{equation}
where we have introduced ${\cal H}_{\{\hat\Pi_i\}}(A|B){\equiv}\sum_{i}p_i{\cal S}(\rho^i_{A|B})$ as the quantum conditional entropy associated with the the post-measurement density matrix $\rho^i_{A|B}=\text{Tr}_{B}[\hat\Pi_i\rho_{AB}]/p_i$ obtained upon performing the complete projective measurement $\{\Pi_i\}$ on qubit $B$. QD is thus defined as
\begin{equation}
{\cal D}^\leftarrow\,{=}\,\inf_{\{\Pi_i\}}[{\cal I}(\rho_{AB})-{\cal J}^\leftarrow(\varrho_{AB})]
\end{equation}
with the infimum calculated over the set of projectors $\{\hat\Pi_i\}$ \cite{discord1,paternostro}. Analogously, one can define ${\cal D}^\rightarrow$, which is obtained upon swapping the roles of $A$ and $B$.

On the other hand, our chosen entanglement measure is EoF~\cite{wootters}, which quantifies the minimum number of Bell pairs needed in order to prepare a copy of the state we are studying. For arbitrary two-qubit states, EoF is calculated as
\begin{equation}
\mathcal{E}\,{=}\,h\left(\frac{1}{2}\left[1+\sqrt{1-C_{in}^2}\right]\right)
\end{equation}
where $h(x)\,{=}\,-x\text{log}_2 x - (1-x)\text{log}_2 (1-x)$ is the binary entropy function and $C_{in}$ is the concurrence of the state~\cite{wootters}. The latter, an equally valid entanglement measure, is found in terms of the eigenvalues $\lambda_1\,{\geq}\, \lambda_{2,3,4}$ of the matrix $\rho_{AB}(\hat\sigma_y\otimes\hat\sigma_y)\rho^*_{AB} (\hat\sigma_y\otimes\hat\sigma_y)$ as
\begin{equation}
C_{in}=\text{max}\left[0,\sqrt{\lambda_1}-\sum_{i=2}^{4}\sqrt{\lambda_i}\right],
\end{equation}
where $\hat\sigma_y$ is the $y$-Pauli operator.

Comparing two measures of quantum correlation stands as fundamental problem. Even if we consider the same figure of merit, e.g. entanglement, we are plagued by ordering problems between different measures when dealing with an arbitrary (mixed) state. Such problems become compounded if one attempts to draw conclusions between two different figures of merit, such as entanglement and discord, for an arbitrarily chosen state. Therefore we will restrict our study to initially pure states of the form
\begin{eqnarray}
\label{initialphi}
\ket{\Phi}&=&\alpha \ket{00} + \sqrt{1-\alpha^{2}} \ket{11}\\
\label{initialpsi}
\ket{\Psi}&=&\alpha \ket{01} + \sqrt{1-\alpha^{2}} \ket{10}
\end{eqnarray}
where we have parameterized the amount of initial correlation in the state in terms of the concurrence $C_{in}$ as
\begin{equation}
\alpha=\sqrt{\frac{1+\sqrt{1-C_{in}^2}}{2}},
\end{equation} 
Given $C_{in}\!\in\![0,1]$ requires $1/2\leq\alpha\leq1$. As mentioned, when the state at hand is pure, the EoF and QD are exactly equivalent, both quantitatively and from an operational view point, and are given by the von Neumann entropy. This will allow us to have a justifiable starting point from which we will be able to compare the measures even when the state has become mixed due to the interaction with the environments. 

\begin{figure}[t]
{\bf (a)} \hskip4cm {\bf (b)}
\psfig{figure=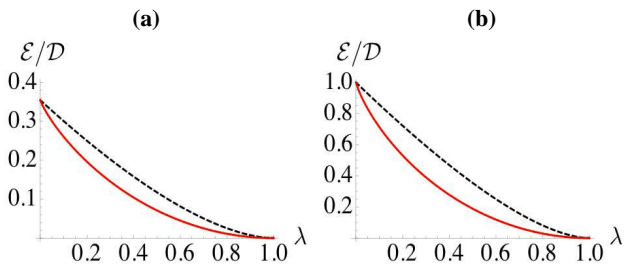,width=8.5cm,height=3.2cm}
\caption{EoF black (dashed) and QD red (solid) for a phase damping affected pair of initially pure qubits. Each plot is for an increasing amount of initial correlation corresponding to a $C_{in}=$ {\bf (a)} 0.5, and {\bf (b)} 1.0.}
\label{phase1}
\end{figure}
\section{Dynamics of quantum Correlations }
\subsection{Phase Damping}
\label{phasedamping}
Phase damping provides a fertile area to study the behavior of quantum correlations since the nature of the noise is one which only affects the quantum coherences and will leave the classical correlations intact. Through Eqs. (1) and (3) we can readily find the damped density matrix and explore the how the different construction of correlation quantifiers behave under the same type of noisy channel. A further remark is in order: it is immaterial which form of initial state Eq.~(\ref{initialphi}) or (\ref{initialpsi}) we consider. This is a consequence of the fact the channel only effects  the off diagonal terms appearing in the density matrix and has no affect on the populations. Hence it does not matter which subspace of the Hilbert space we are confined to.

By examining Fig.~\ref{phase1} we find that regardless of the initial correlations and of the strength of the channel, QD decreases at a faster rate than EoF. This would appear counterintuitive considering the broad definition of the the two figures of merit, QD by nature includes states that are entangled, and indeed one cannot have an entangled state with zero QD. Thus the natural answer to such a counterintuitive result is to conclude that the two measures are fundamentally incomparable. However, given the carefully chosen initial states and figures of merit, we see that the two measures are identical in their definition initially. 
By taking these conditions into account one must reach the conclusion that quantitatively EoF performs better than QD under phase damping.

In order to explain why this is the case we scale the problem back so we can make use of the results of Ref.~\cite{fanchini}. If we consider an asymmetric damping, such that only one qubit in the initially correlated pair is damped and without loss of generality assuming Eq. (4) as our initial state the total system-environment state can be written
\begin{equation}
\label{pure}
\ket{\psi}_{ABE}=\alpha\ket{000}+\sqrt{1-\alpha^2}\left(\sqrt{1-\lambda}\ket{101}+\sqrt{\lambda}\ket{111}\right).
\end{equation}
This state is a pure tripartite entangled state. In Ref.~\cite{fanchini} the authors proved a strict conservation relation between the EoF and QD for any mixed bipartite states by examining the distribution of the correlations in a tripartite pure state. As stated in~\cite{fanchini} 
\begin{quote}
{\it ``Given an arbitrary tripartite pure system, the sum of all possible bipartite entanglement shared with a particular subsystem, as given by the EoF, cannot be increased without increasing, by the same amount, the sum of all QD shared with this same subsystem."} 
\end{quote}
Hence,
\begin{equation}
\label{conservation}
\mathcal{E}_{AB}+\mathcal{E}_{AE}=\mathcal{D}_{AB}^\gets+\mathcal{D}_{AE}^\gets.
\end{equation}
The asymmetry usually present when dealing with QD is not an issue for us as under this noise QD is symmetrical for our initial states. If we now examine what happens for Eq.~(\ref{pure}) we find that while the damped qubit $A$ develops a non-zero QD with the environment $E$, the pair never share any entanglement, i.e. $\mathcal{E}_{AE}=0~\forall~\lambda,\alpha$. Hence 
\begin{equation}
\mathcal{E}_{AB}=\mathcal{D}_{AB}+\mathcal{D}_{AE},
\end{equation}
therefore QD, in this case, can never be larger than EoF. For the case of only one of the qubits damped this explains the better resilience of EoF to phase 
damping over QD. As $\lambda\to1$ we see all quantum correlations are lost and we are left with a completely classical total state of qubit-environment. 

\begin{figure}[t]
{\bf (a)} \hskip5cm {\bf (b)} \hskip5cm\\
{\includegraphics[width=4.5cm]{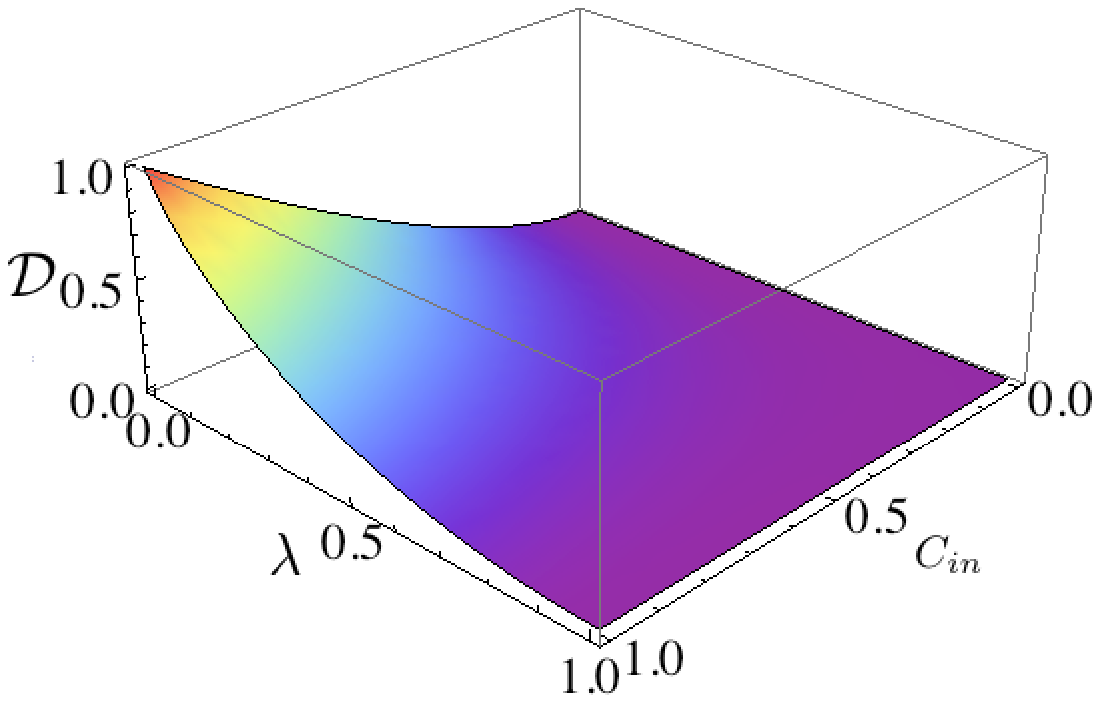}}~~{\includegraphics[width=4.5cm]{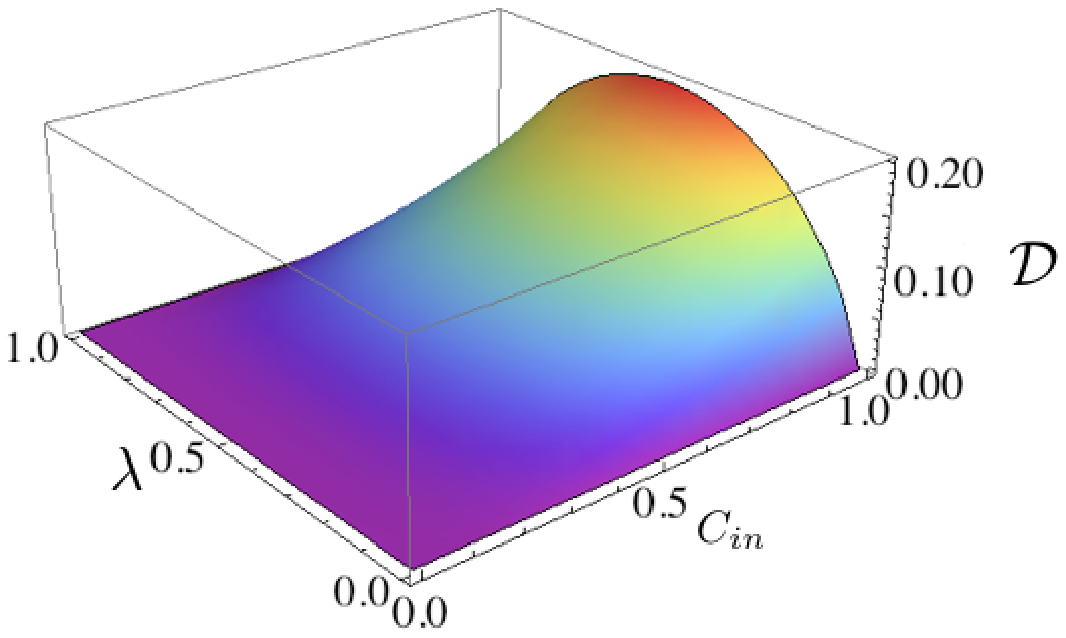}}\\
{\bf (c)}\\
{\includegraphics[width=4.5cm]{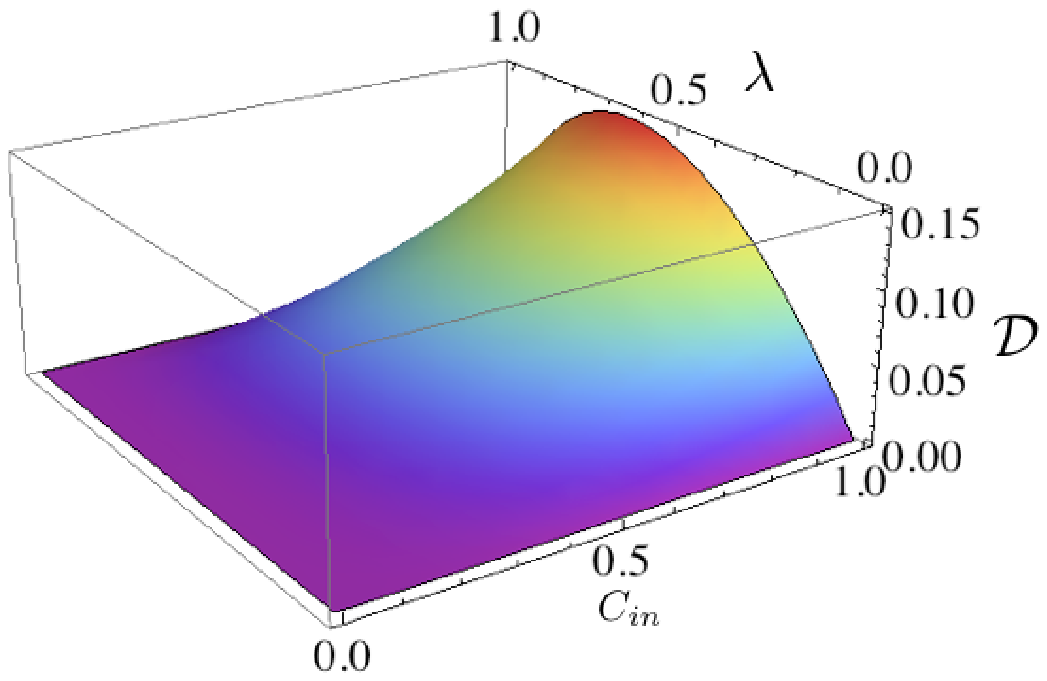}}
\caption{Quantum discord of {\bf (a)} $\rho_{AB}$, {\bf (b)} $\rho_{AE_{A}}$ and  {\bf (c)} $\rho_{E_A E_B}$. Entanglement is zero everywhere except in $\rho_{AB}$. Plots for phase damping affecting both qubits. See text for discussion}
\label{phase2}
\end{figure}

\begin{figure*}[t]
{\bf (a)} \hskip4cm {\bf (b)} \hskip4cm {\bf (c)}\\
{\includegraphics[width=5.5cm]{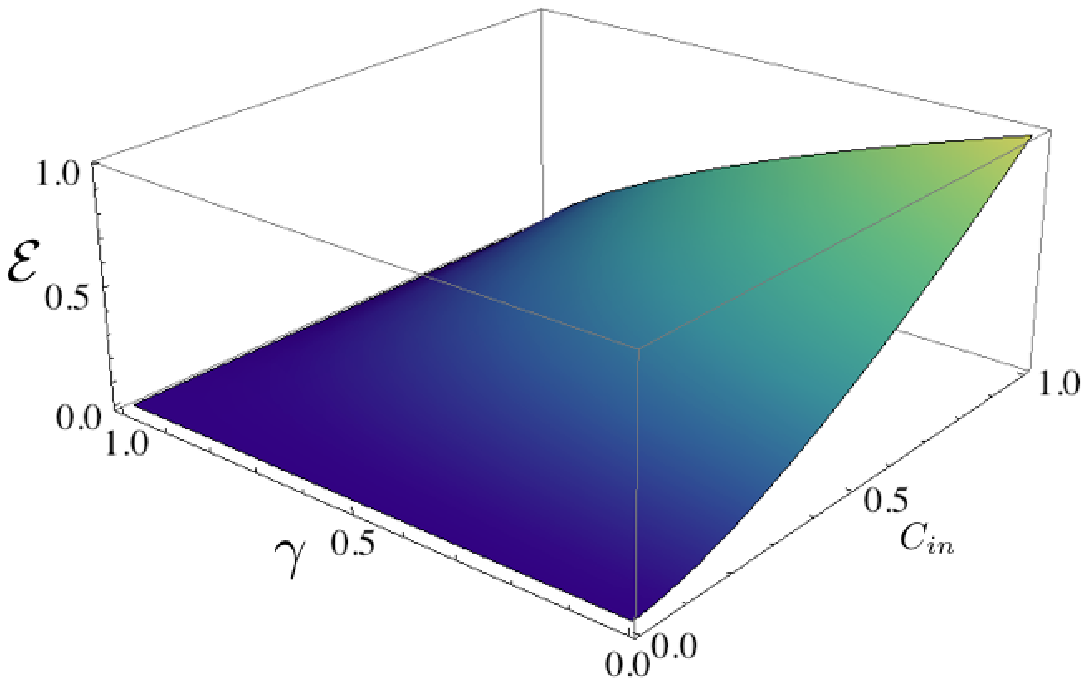}}~~{\includegraphics[width=5.5cm]{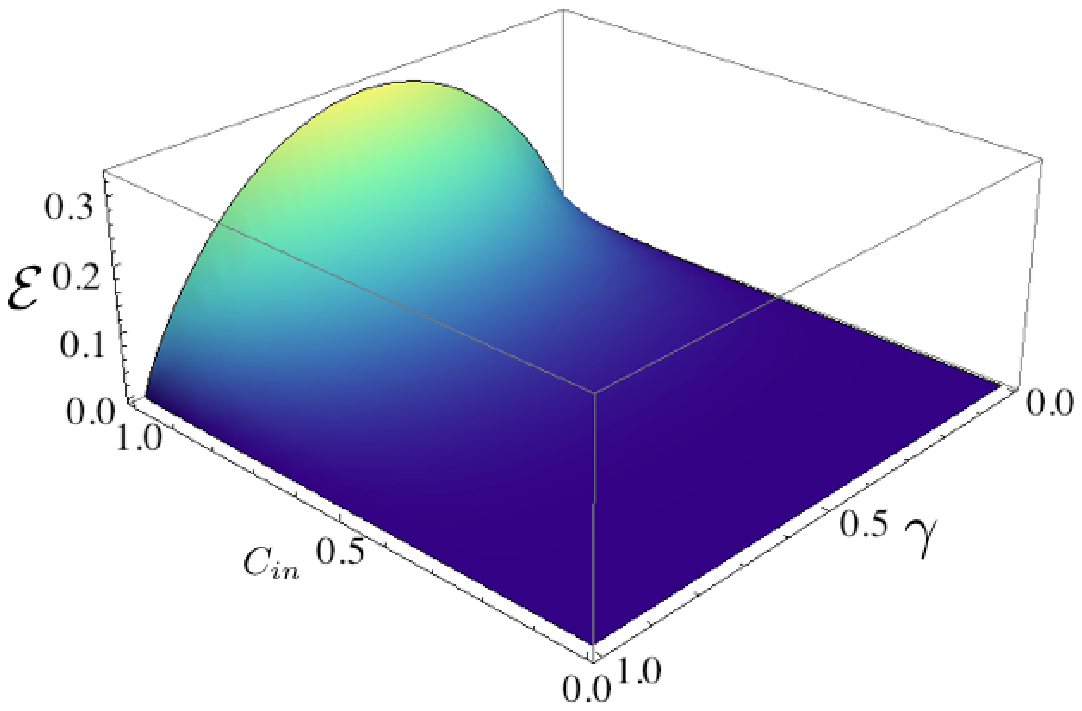}}~~{\includegraphics[width=5.5cm]{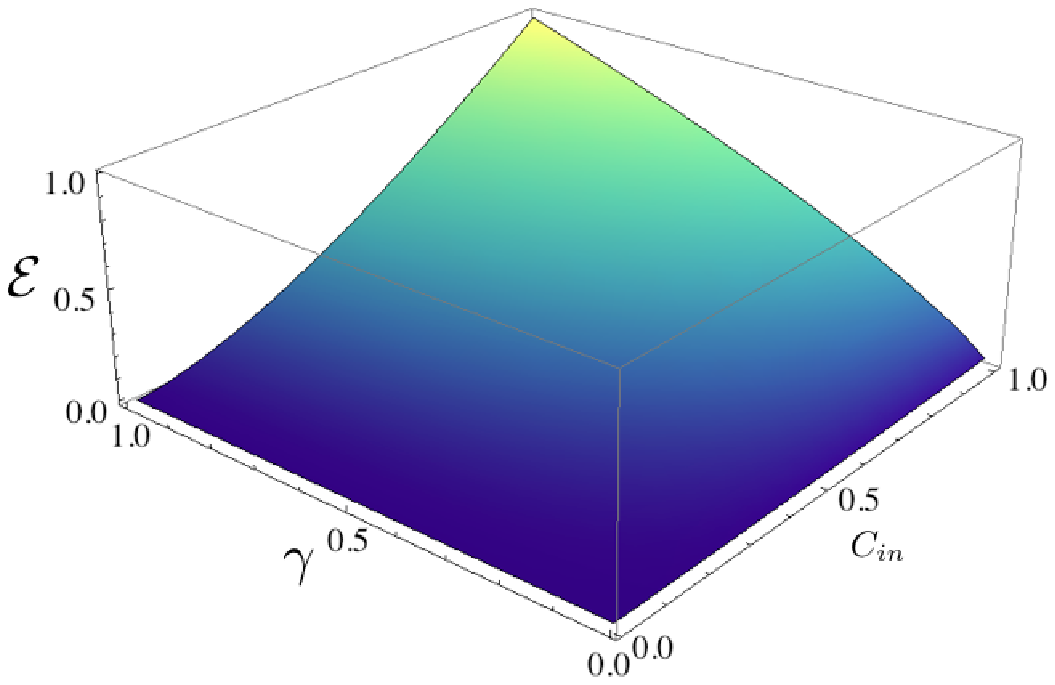}}\\
{\bf (d)} \hskip4cm {\bf (e)} \hskip4cm {\bf (f)}\\
{\includegraphics[width=5.5cm]{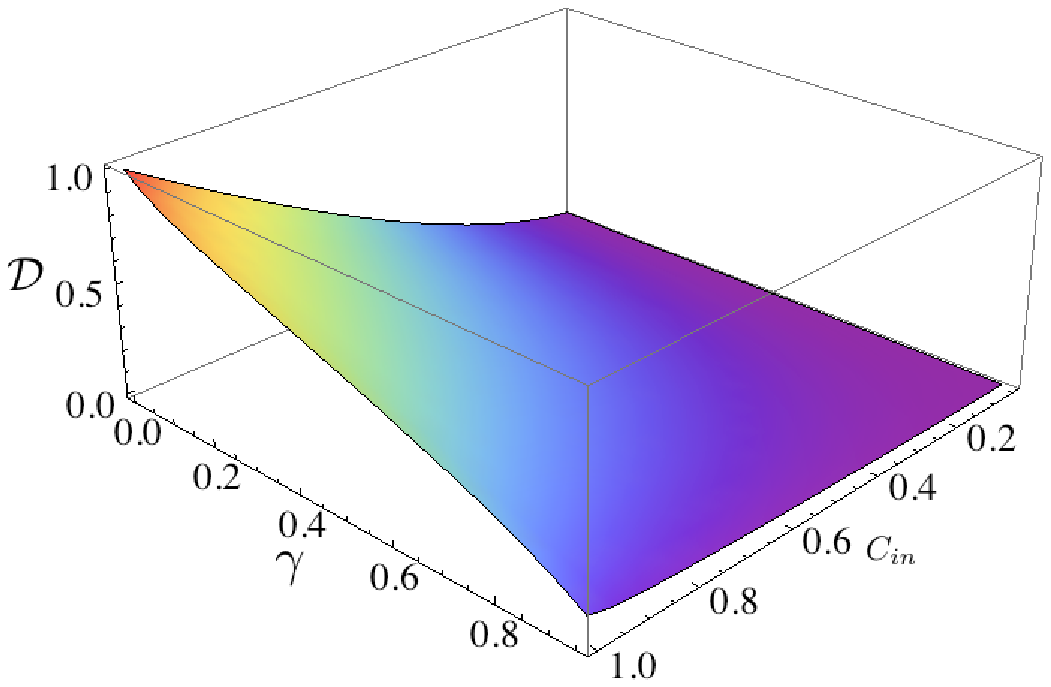}}~~{\includegraphics[width=5.5cm]{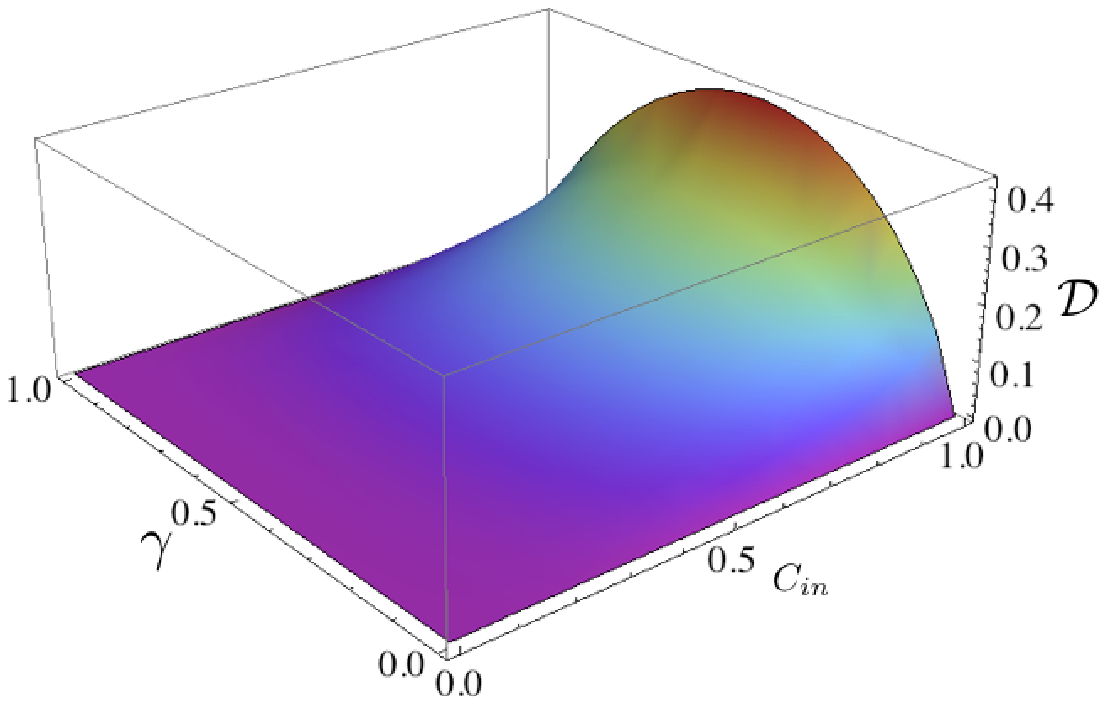}}~~{\includegraphics[width=5.5cm]{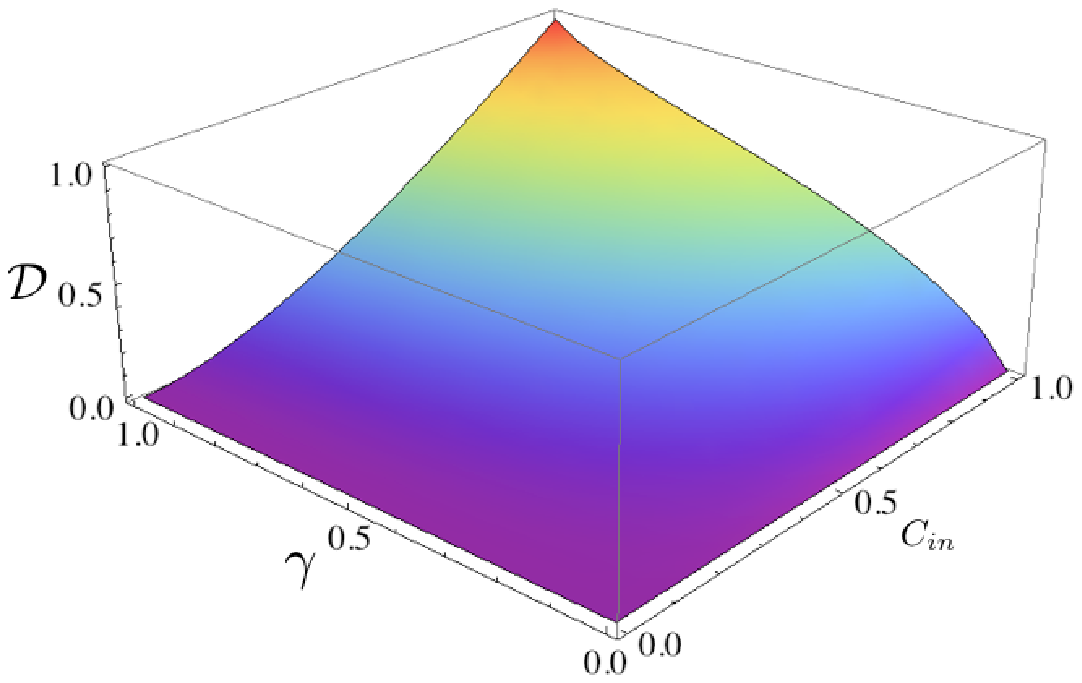}}
\caption{EoF of {\bf (a)} $\rho_{AB}$, {\bf (b)} $\rho_{AE}$ and {\bf (c)} $\rho_{BE}$ and one-way QD of {\bf (d)} $\rho_{AB}$, {\bf (e)} $\rho_{AE}$ and {\bf (f)} $\rho_{BE}$ for amplitude  damping affecting only qubit $A$. See text for discussion.}
\label{amp1}
\end{figure*}

When both qubits are damped the system exhibits an analogous behavior. However we are unable to exploit any strict relations like Eq.~(\ref{conservation}). Also as shown in~\cite{giorgi,prabhu} EoF and QD do not, in general, exhibit monogamic behavior. In Fig.~\ref{phase2} we plot the QD for $\rho_{AB}$, $\rho_{AE_A}$,
and $\rho_{E_A E_B}$. Here it is clear as the correlation shared between the two qubits is reduced due to their interaction with the environments there is a non-zero QD developed between all the other bipartitions of the composite system. However, with the exception of $\rho_{AB}$, the entanglement in all of these bipartitions is exactly zero and thus the entanglement is quantitatively and qualitatively larger than discord. Although for ease of calculation we have assumed the damping rate in each environment is the same, such a behavior evidently holds even when the damping parameters for the environments are different.

\subsection{Amplitude Damping}
\label{ampdamping}
We now consider the amplitude damping channel given by Eq.~(\ref{amp}). If we first consider the case in which only a single qubit is damped we can once again use the results of~\cite{fanchini} to contrast quantitatively EoF and QD although care must be taken. The form of the initial state can now have an impact on the dynamics of quantum correlations due to the changing energy within the system. The states resulting from the action of the channel on qubit $A$ only is
\begin{eqnarray}
\label{phiamp}
\ket{\psi_{\Phi}}_{ABE}=\alpha\ket{000}+\sqrt{1-\alpha^2}\left(\sqrt{\gamma}\ket{011}+\sqrt{1-\gamma}\ket{110}\right),\\
\label{psiamp}
\ket{\psi_{\Psi}}_{ABE}=\alpha\ket{010}+\sqrt{1-\alpha^2}\left(\sqrt{\gamma}\ket{001}+\sqrt{1-\gamma}\ket{100}\right).
\end{eqnarray}
Due to the asymmetry of QD now we must be strict in the subsystem that we perform the measurements on when calculating the QD in order to make use of the conservation relation Eq.~(\ref{conservation}). If we restrict to only measurement on $A$ then the reduced state dynamics for all pairs will be independent of what the  initial state was. This is simply because by restricting to measurements on $A$ the formula for QD does not take into account the differences arising by starting from an initial state in the single- or two-excitation subspace. 

The qualitative behavior of the system is very different under amplitude damping. There is clearly an energy exchange happening between the system and environment and therefore it is intuitive to expect some build of correlations between the system and environment. And, in fact, when the qubit $A$ is maximally damped all correlations between $A$ and $B$ are lost. However we find the reduced state $\rho_{BE}$ now shares precisely the same amount of initial correlations. Hence the amplitude damping admits an effective entanglement swapping action whereby the correlations are exchanged from $A$ to the environment. This is in contrast to the phase damping case address previously, since there only coherences were diminished and there is no exchange of correlations when the system is maximally damped. In Fig.~\ref{amp1} we plot the EoF and one-way QD, measured on the damped qubit $A$, between the various reduced states of Eq.~(\ref{phiamp}). The first major difference we notice is $\mathcal{E}_{AE}\neq0$. As Eq.~(\ref{phiamp}) is pure, and making use of the conservation relation Eq.~(\ref{conservation}), we find regardless of $\gamma$ or $\alpha$
\begin{equation}
\mathcal{E}_{AE}\leq\mathcal{D}_{AE}^\gets,
\end{equation}
with equality only holding when $\gamma=0$ or 1. Hence by the conservation relation the EoF in $\rho_{AB}$  will always be greater than the QD. In this case we again see entanglement serves as a more robust form of quantum correlations over discord. 

\begin{figure*}[t]
{\bf (a)} \hskip5cm {\bf (b)} \hskip5cm {\bf (c)}
\psfig{figure=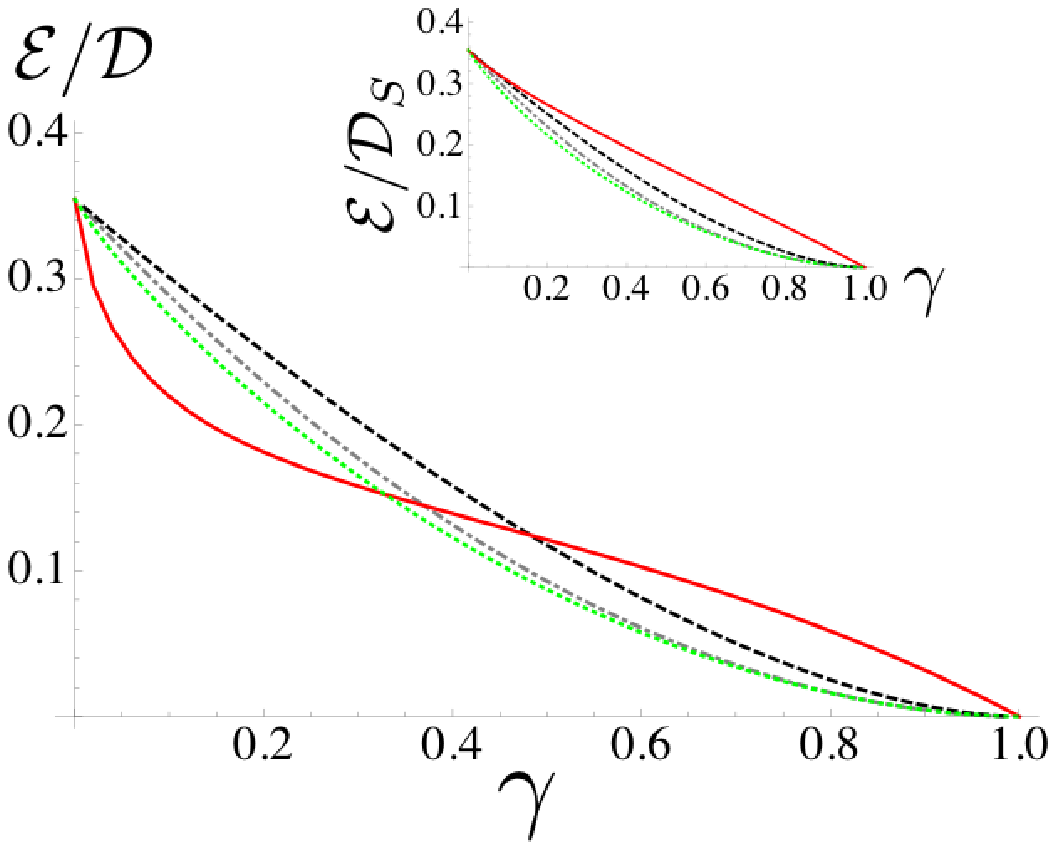,width=6cm,height=4.5cm}~~\psfig{figure=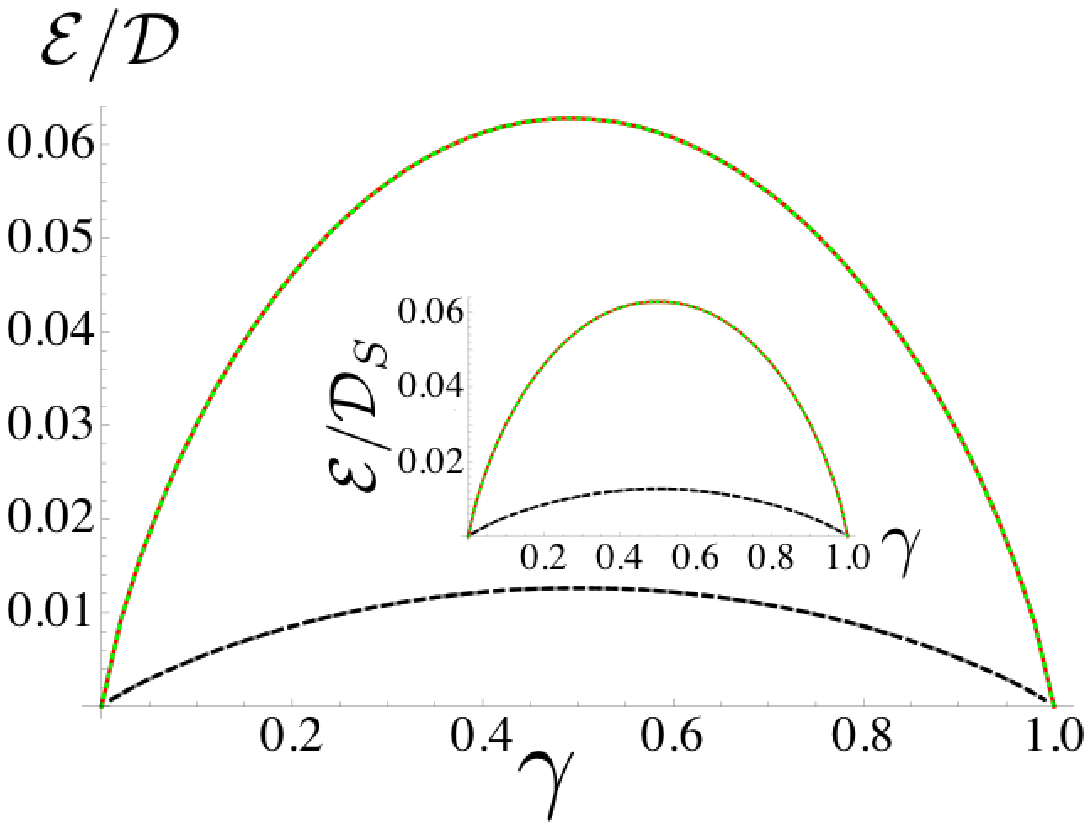,width=6cm,height=4.5cm}~~\psfig{figure=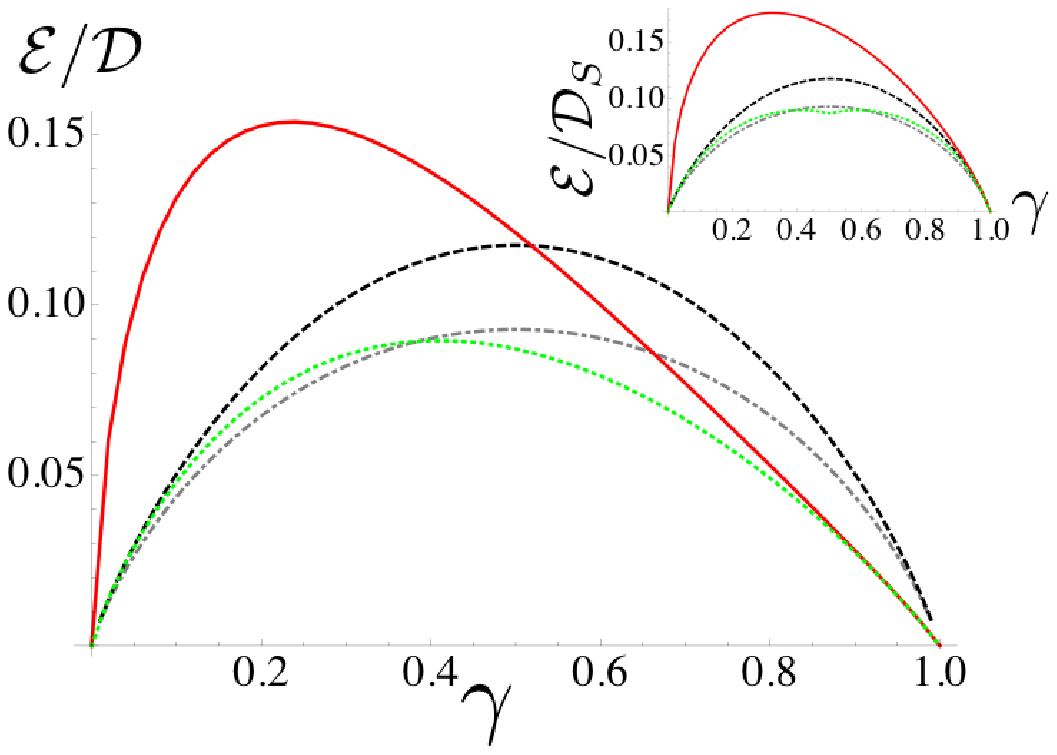,width=6.2cm,height=4.5cm}
\caption{EoF (dashed, black), one-way QD (solid, red) for 2-qubits initially in $\ket{\Psi}$-state, Eq.~(\ref{initialpsi}), and EoF (gray, dot-dashed), one-way QD (green, dotted) for 2-qubits initially in $\ket{\Phi}$-state, Eq.~(\ref{initialphi}) both affected by amplitude damping channels. All insets show the symmetrized version of QD Eq.~(\ref{symdiscord}) for the same states. In all plots $C_{in}$=0.5.}
\label{amp2}
\end{figure*}

If we consider a symmetrized version of QD, sometimes referred to as ``two-way" QD~\cite{paternostro},
\begin{equation}
\label{symdiscord}
\mathcal{D}_S=\text{Max}[\mathcal{D}^\to,\mathcal{D}^\gets],
\end{equation}
Eq.~(\ref{conservation}) no longer holds. Although EoF {\it typically} performs better than QD there are instances when the initial state has $C_{in}\simeq1$ when QD is the better conserved quantity. However, making definitive statements comparing the two measures is dangerous. Entanglement by construction makes no distinction between the subsystems within the state while discord does. The asymmetry of QD poses a significant issue if we wish to make quantitative comparisons. We are perfectly justified in concluding EoF is a preferred quantity when we consider amplitude damping if we measure only on qubit $A$, while the ``two-way" QD cannot be used to compare the measures since it is no longer fully rooted in the same operational definition as EoF due to the maximization in Eq.~(\ref{symdiscord}).

When both qubits are damped these issues become compounded significantly, as shown in Fig.~\ref{amp2}. When both qubits are damped, both EoF and QD are now sensitive to the form of the initial state. This means that any firm conclusions on the comparative behavior is difficult. Clearly there are regions in which one measure appears preferable to another, but with the lack of any strict relation there is little meaning to the comparison. 

\section{Mixed initial states}
\label{mixedstates}
For completeness we now turn our attention to mixed initial states. Exploring the behavior for a general two qubit mixed state is computationally extremely costly and provides little insight to our study. To reduce the calculation complexity we examine what happens when the initial state of the two qubits is in a Werner state~\cite{wernerref}
\begin{equation}
\label{werner}
\varrho_W=\eta \ket{\psi}\bra{\psi} + \frac{(1-\eta)}{4}\openone.
\end{equation}
where $\ket{\psi}$ is a maximally entangled pure Bell state and $0\leq\eta\leq1$. As our starting state is no longer pure we find EoF and QD do not coincide initially. In fact, Eq.~(\ref{werner}) is separable for $\eta\leq1/3$ while it has a non-zero discord for the whole range of $\eta\!\in\!(0,1]$. In order to attempt to remove this discrepancy we shall focus on a rescaled quantity for the correlations, i.e. $\mathcal{E}/\mathcal{E}_{in}$ ($\mathcal{D}/\mathcal{D}_{in}$), where $\mathcal{E}_{in}$ ($\mathcal{D}_{in}$) is the initial value of EoF (QD) for the undamped system. The result of this means we are asking what portion of correlations, as quantified by a given figure of merit, are degraded by the noisy environment. 

\begin{figure}[b]
{\bf (a)} \hskip3.5cm {\bf (b)}
\psfig{figure=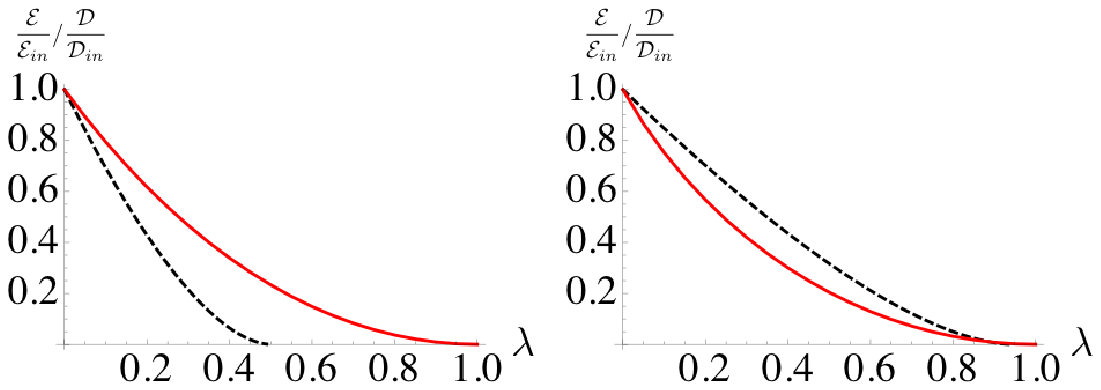,width=9cm,height=3.2cm}\\
{\bf (c)} \hskip3.5cm {\bf (d)}\\
\psfig{figure=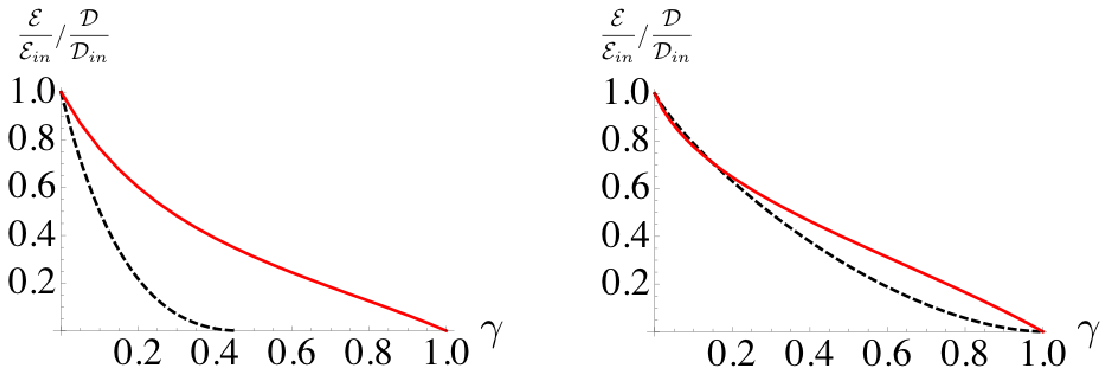,width=9cm,height=3.2cm}
\caption{Rescaled EoF (dashed, black) and rescaled QD (solid, red) for the 2-qubit mixed state Eq.~(\ref{werner}) under going {\bf (a),~(b)} phase damping and {\bf (c),~(d)} amplitude damping. $\eta=0.5$ panels {\bf (a)} and {\bf (c)}, $\eta=0.95$ panels {\bf (b)} and {\bf (d)}.}
\label{wernerfig}
\end{figure}

Fig.~\ref{wernerfig} shows the behavior of QD and EoF of Eq.~(\ref{werner}) for phase damping, panels {\bf (a)} and {\bf (b)}, and amplitude damping, panels {\bf (c)} and {\bf (d)}. In the case of phase damping if the initial correlation is not large enough we find the rescaled QD appears more robust to noise than EoF, however the larger the initial correlation, and hence the less mixed the initial state, we see this behavior reversed. Amplitude damping behaves quite differently. Here the rescaled QD is almost always outperforming EoF, until the state is nearly pure. This behavior is explained in~\cite{campbell} where the ability to propagate nonclassical correlations was explicitly linked to the probability of exchanging an excitation.

Clearly, quantitatively comparing EoF and QD here is dangerous. Despite taking a rescaled quantity in an attempt to make the measures comparable, the fact that QD and EoF do not coincide initially leads to ambiguous results. Before we can make a definitive conclusion a deeper understanding of the interplay between entanglement and discord for mixed states is needed. 

\section{Conclusions and Discussions}
\label{conclusions}
In this paper we have examined how both quantitatively and qualitatively different figures of merit for quantum correlations behave under noisy processes. In order to make such comparisons we considered two correlation measures that share the same entropic definitions, namely entanglement of formation (EoF) and quantum discord (QD). These measures are such that when the state is pure, both are equivalent and equal to the von Neumann entropy. We considered a broad class of states that are pure and parameterized by their concurrence. The quantum channels we examined were the Markovian phase and amplitude damping channels.  Under these circumstances we see it can be meaningful to make both qualitative and quantitative comparisons between the two measures of quantum correlation. Interestingly we have shown the EoF serves as a preferred resource over QD when the noisy process affects only the quantum correlations and leaves classical correlations and the energies unaffected. By exploiting the strict relation between EoF and QD derived in~\cite{fanchini} we were able to show that the reason for EoFs better performance was due to the fact that at no point during the dynamical evolution did the damped system ever develop entanglement with the environment, while there was a non-zero QD developed. When the noisy process was one such that the exchange of excitations between the system and environment was allowed, drawing firm conclusions was more difficult. The results highlighted another issue arising from the asymmetry of QD. When exploring the effects of amplitude damping, due the exchange of excitations with the environment the QD was sensitive to the form of the initial state and on which subsystem we measured. If we consider only one-way QD and only a one qubit of the pair damped, then by the same reasoning described for phase damping we can see the EoF is a more robust quantifier of quantum correlations. However, considering the symmetrized QD we can no longer be so conclusive. And when we considered both qubits damped we lack any tools to make a meaningful comparison. The results of~\cite{cornelio}, concerning the emergence of the pointer basis in the dynamics of correlations, offer some additional insights into the results presented here.

Making comparisons between measures has stood as a difficult task in quantum theory. Here we consider some necessary constraints in order to quantitatively compare the measures: the two measures are operationally equivalent and, must coincide initially. In~\cite{adesso} a strict relationship between negativity and the geometric measure of quantum discord was proven for all two qubit states, $\mathcal{N}^2\leq\mathcal{G}_D$. If we consider the exact same analysis performed in Sec.~\ref{phasedamping} we find this inequality is saturated. However, examining the dynamics of the reduced states of $\rho_{AE}$ we still find a non-zero geometric discord, while negativity is exactly zero. This highlights some recent concerns associated with the geometric discord~\cite{piani}. As was clearly shown in~\cite{adesso} mathematically comparing these measures is well justified, however such a comparison lacks a physical motivation. It may be tempting to rely on other distance based measures such as the relative entropy of entanglement ($E_R$) and discord ($D_R$). However, in this case $D_R$ is likely to be unfairly favored since the set of classical states is a subset of the set of separable states, i.e. a state cannot be classical and non-separable, while the converse is easily possible. The difficultly in attempting to use these measures is more evident if we consider a phase damping affected pair of qubits. Since the noise only affects the coherences it is fairly easy to see that the closest classical state is exactly the closest separable state resulting in $E_R=D_R$, however there is still QD developed between the environment and system while entanglement is exactly zero. The same situation encountered when considering squared negativity and geometric discord. If we remove the condition for the states to be initially pure, and instead consider the Werner states as our initial states, then once again we lose the ability to make firm conclusions. This is because the two measures are no longer equal due to the initial mixedness. 

The results raise a number of important questions, principally: when can we compare entanglement and discord? Here we made a physically motivated approach and explored the measures behavior under noise. The conditions considered allowed us to show how the conservation relation Eq.~(\ref{conservation}) can be used to make a comparative picture of quantum correlations. Also, some studies have shown QD to perform better than entanglement, e.g. ~\cite{bruss,campbell,francesco}, something that would appear intuitively normal given the definition of the quantifiers, while here we show the converse can also be true. Indeed, if we view quantum correlations as potential information processing resources then it is important to understand what is the advantage of exploiting one form of correlation over the other. The sole resilience to noise does not imply advantages of one potential resource over the others. While discord has been shown to be an attractive candidate for information processing~\cite{datta,arxiv} our results highlight the need for a deeper understanding of the nature of quantum correlations at a fundamental level. In addition, as shown in~\cite{fanchinichina}, this predominance of EoF over QD can extend to multipartite systems and the authors reached some complimentary conclusions to those presented here. We expect our results will serve as an initial step to determining a broader framework for the understanding and comparison between different figures of merit for quantum correlations.

\acknowledgements
The author greatly acknowledges fruitful discussions and exchanges with Drs. Mauro Paternostro, Tomasz Paterek, Laura Mazzola, Thomas Busch, Gianluca Giorgi and Dave Rea.
%\begin{figure}[h!]
%\includegraphics[scale=0.29]{kari1.eps}~\includegraphics[scale=0.26]{kari2.eps}\\
%\end{figure}

\end{document}